\documentstyle[latexsym,multicol,aps,prl,epsfig]{revtex}
\renewcommand{\narrowtext}{\begin{multicols}{2} \global\columnwidth20.5pc}
\begin{document}

\title{Limit on Lorentz and CPT Violation of the Proton Using a
  Hydrogen Maser}
 
\author{D.\ F.\ Phillips, M.\ A.\ Humphrey, E.\ M.\ Mattison, \\
R.\ E.\ Stoner, R.\ F.\ C.\  Vessot and R.\ L.\ Walsworth}
\address{Harvard--Smithsonian Center for Astrophysics,  Cambridge, MA 02138}

\date{submitted to \emph{Physical Review Letters} on August 27, 2000}

\maketitle 
 
\begin{abstract}
  We present a new measurement constraining Lorentz and CPT violation
  of the proton using a hydrogen maser double resonance technique.  A
  search for hydrogen Zeeman frequency variations with a period of the
  sidereal day (23.93 h) sets a clean limit on violation of Lorentz
  and CPT symmetry of the proton at the $10^{-27}$ GeV level.
\end{abstract}

\narrowtext 

Experimental investigations of Lorentz symmetry provide important
tests of the standard model of particle physics and general
relativity.  While the standard model successfully describes particle
phenomenology, it is believed to be the low energy limit of a
fundamental theory that incorporates gravity.  This underlying theory
may be Lorentz invariant, yet contain spontaneous symmetry-breaking
that could result at the level of the standard model in small
violations of Lorentz invariance and CPT (symmetry under simultaneous
application of Charge conjugation, Parity inversion, and Time
reversal).

Clock comparisons \cite{clockexperiments,hl9x} provide sensitive tests
of rotation invariance and hence Lorentz symmetry by bounding the
frequency variation of a given clock as its orientation changes, e.g.,
with respect to the inertial reference frame defined by the fixed
stars \cite{clockcompare}.  Atomic clocks are typically used,
involving the electromagnetic signals emitted or absorbed on hyperfine
or Zeeman transitions. Here we report results from a hydrogen (H)
maser experiment that sets an improved clean limit on Lorentz and CPT
violation of the proton at the level of $10^{-27}$ GeV as the H maser
rotates with the Earth.

Our H maser measurement is motivated by a standard model extension
developed by Kosteleck\'y and
others~\cite{clockcompare,spontlorentz,stringLorentz,kosanalysis,bkrhydrogen}.
This standard-model extension is quite general: it emerges as the
low-energy limit of any underlying theory that generates the standard
model and that contains spontaneous Lorentz symmetry violation
\cite{spontlorentz}.  For example, such characteristics might emerge
from string theory \cite{stringLorentz}.  A key feature of the
standard-model extension is that it is formulated at the level of the
known elementary particles, and thus enables quantitative comparison
of a wide array of searches for Lorentz and CPT violation
\cite{kosanalysis}.
The dimensionless suppression factor for
such effects would likely be the ratio of the appropriate low-energy
scale to the Planck scale, perhaps combined with dimensionless
coupling
constants~\cite{clockcompare,spontlorentz,stringLorentz,kosanalysis,bkrhydrogen}.

Recent experimental work motivated by this standard-model extension
includes Penning trap tests by Gabrielse \emph{et al.} on the
antiproton and H$^-$ \cite{gab99}, and by Dehmelt \emph{et al.} on the
electron and positron \cite{deh99}, which place improved
limits on Lorentz and CPT violation in these systems.  A re-analysis by
Adelberger, Gundlach, Heckel, and co-workers of existing data from the
``E\"ot-Wash II'' spin-polarized torsion pendulum
\cite{adelharris} sets the most stringent bound to date on Lorentz
and CPT violation of the electron: approximately $10^{-29}$ GeV
\cite{hec99}. A recent search for Zeeman-frequency sidereal variations
in a ${}^{129}$Xe/${}^3$He maser places an improved constraint on Lorentz
and CPT violation involving the neutron at the level of $10^{-31}$ GeV
\cite{DNGMlorentz}. Also the KTeV experiment at Fermilab and the OPAL and
DELPHI collaborations at CERN have limited possible Lorentz and CPT
violation in the $K$ and $B_d$ systems \cite{kbexpt}.

\begin{figure}
\centerline{\epsfig{figure=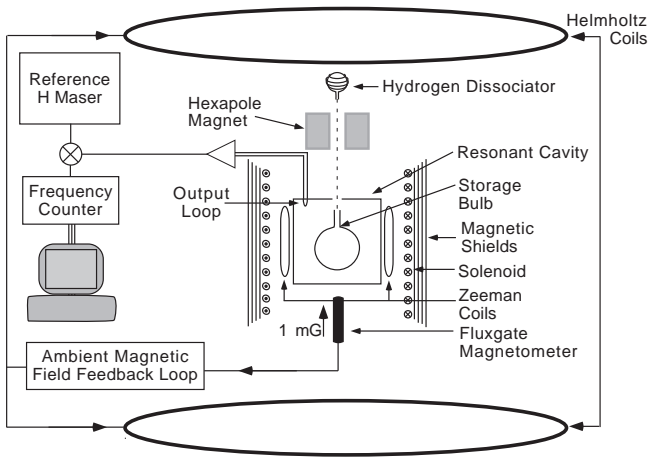,width=\hsize,clip=}} 
\smallskip
\caption{Schematic of the H maser in its ambient magnetic field
  stabilization loop.  Large Helmholtz coils surround the maser and
  cancel external field fluctuations as detected by a fluxgate
  magnetometer placed close to the maser region. Zeeman coils mix the
  $m_F$ sublevels of the $F=1$ hyperfine state, and allow sensitive
  measurement of the Zeeman frequency through pulling of the maser
  frequency \protect\cite{andresen}, as determined by
  comparison to a reference H maser.}
\label{f.Hschem}
\end{figure}

The hydrogen maser is an established tool in precision tests of
fundamental physics \cite{happs}. H masers operate on the $\Delta F =
1$, $\Delta m_F = 0$ hyperfine transition (the ``clock'' transition)
in the ground electronic state of atomic hydrogen \cite{kgrvv}.
Hydrogen molecules are dissociated into atoms in an RF discharge and
the atoms are spatially state selected via a hexapole magnet (Fig.\ 
\ref{f.Hschem}). Atoms in the $F=1$, $m_F = +1,0$ states are focused
into a Teflon coated cell, thereby creating the population inversion
necessary for active maser oscillation.  The cell resides in a
microwave cavity resonant with the $\Delta F=1$ transition at 1420
MHz.  A static magnetic field of $\sim 1$ milligauss is applied by a
solenoid surrounding the resonant cavity to maintain the quantization
axis of the H atoms.  For normal H maser operation, this magnetic
field is directed vertically upwards in the laboratory reference
frame. The $F=1$, $m_F=0$ atoms are stimulated to make a transition to
the $F=0$ state by the thermal microwave field in the cavity.  The
energy from the atoms then acts as a source to increase the microwave
field. With sufficiently high polarization flux and low cavity losses,
this feedback induces active maser oscillation. H masers built in our
laboratory over the last 30 years provide fractional frequency
stability on the clock transition of better than $10^{-14}$ over
averaging intervals of minutes to days and can operate undisturbed
for several years before requiring routine maintenance.

\begin{figure}
\centerline{\epsfig{figure=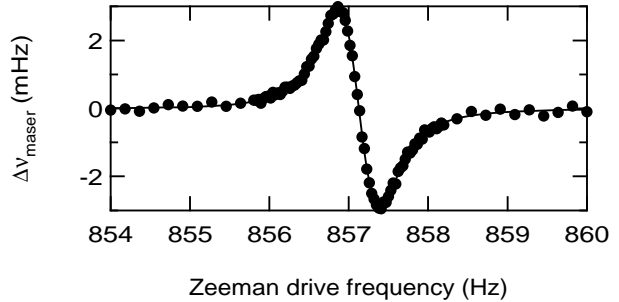,width=\hsize,clip=}} 
\smallskip
\caption{
An example of a double resonance measurement of the  $F=1$,
  $\Delta m_F = \pm 1$ Zeeman frequency ($\nu_Z$) in the H maser. 
The
  change from the unperturbed maser clock frequency is plotted versus
  the driving field frequency. 
(The statistical uncertainty in each
  point is approximately 50 $\mu$Hz.) 
The solid line is the fit of the
  antisymmetric lineshape described in \protect\cite{andresen} to the
  data, yielding $\nu_Z = 857.125 \, \pm \, 0.003$ Hz in this
  example. 
}
\label{f.andresen}
\end{figure}

The $\Delta m_F=0$ clock transition has no leading-order sensitivity
to Lorentz and CPT violation \cite{clockcompare,bkrhydrogen} because
the transition encompasses no change in longitudinal spin orientation.
In contrast, the $F=1$, $\Delta m_F= \pm 1$ Zeeman transitions are
maximally sensitive to potential Lorentz and CPT violation
\cite{bkrhydrogen}.  Therefore, we searched for a Lorentz-violation
signature by monitoring the Zeeman frequency ($\nu_Z \approx 850$ Hz
in a static magnetic field of 0.6 mG) as the laboratory reference
frame rotated sidereally.  We utilized an H maser double resonance
technique \cite{andresen} to measure $\nu_Z$.  We applied a weak,
oscillating magnetic field perpendicular to the static field at a
frequency close to the Zeeman transition, thereby coupling the three
sublevels of the hydrogen $F=1$ manifold \cite{maserology}.  Provided
that a population difference exists between the $m_F= \pm 1$ states,
this coupling alters the energy of the $m_F=0$ state, thus shifting
the measured maser clock frequency in a manner described by a line
shape that is antisymmetric about the Zeeman frequency for
sufficiently small static fields (Fig.\ \ref{f.andresen})
\cite{andresen}. We determined $\nu_Z$ by measuring the resonant
driving field frequency at which the maser clock frequency is equal to
its unperturbed value.  Due to the excellent frequency stability of
the H maser, this double resonance technique allowed the determination
of $\nu_Z$ with a precision of $\sim$ 1 mHz~\cite{zeemanAsym}.

In the small-field limit, the hydrogen Zeeman frequency is
proportional to the static magnetic field.  Four layers of high
permeability magnetic shields surround the maser (Fig.\ 
\ref{f.Hschem}), screening external field fluctuations by a factor of
32,000. Nevertheless, the residual effects of day-night variations in
ambient magnetic noise shifted the measured Zeeman frequency with a 24
hour periodicity which was difficult to distinguish from a true
sidereal (23.93 h period) signal in our data sample.
Therefore, we
employed an active stabilization system to cancel external magnetic
field fluctuations (Fig.\ \ref{f.Hschem}). 
A fluxgate magnetometer sensed the field near the maser cavity with a
shielding factor of only 6 to external magnetic fields due to its
location at the edge of the shields.
A feedback loop controlled
the current in large Helmholtz coils (2.4 m dia.) surrounding the
maser to maintain a constant field.
This feedback loop effectively reduced the sidereal fluctuations of
$\nu_Z$ caused by external fields at the location of the magnetometer
to below 1 $\mu$Hz.

We accumulated data in three separate runs of 11, 9 and 12 days over
the period Nov., 1999 to Mar., 2000.  During data taking, the maser
remained in a closed, temperature controlled room to reduce potential
systematics from thermal drifts that might have 24 hour periodicities.
Each $\nu_Z$ measurement required approximately 20 minutes of data
(Fig.\ \ref{f.andresen}).  We also monitored the H maser amplitude,
residual magnetic field fluctuations, maser and room temperatures, and
the current through the maser solenoid (which set the static magnetic
field).  During the second and third runs, we reversed the direction
of the static magnetic field created by the maser's internal solenoid
in order to investigate possible systematic dependence of the diurnal
variation of $\nu_Z$ on field direction. (No such dependence was
observed.) In the field-reversed configuration, the axial magnetic
field in the storage bulb was anti-parallel to the field near the exit
from the state-selecting hexapole magnet. Thus H atoms traversed a
region of magnetic field inversion on their way into the storage bulb,
causing loss of atoms from the maser excited state ($F=1$, $m_F=0$)
due to Majorana transitions as well as sudden transitions of atoms
from the $F=1$, $m_F=+1$ state to the $F=1$, $m_F=-1$ state.  In the
field reversed configuration, the maser amplitude was reduced by 30\%
and both the maser clock frequency and Zeeman frequency were less
stable. Thus, our constraint on sidereal-period $\nu_Z$ variations was
5 times weaker in the field-reversed configuration than in the
parallel-field configuration.

\begin{figure}
\centerline{\epsfig{figure=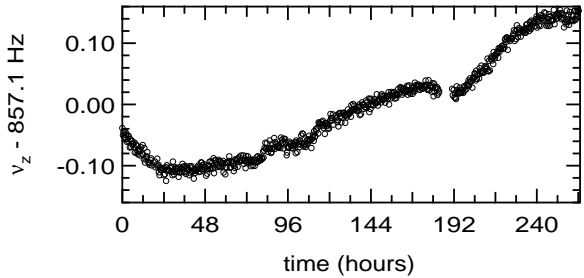,width=\hsize,clip=}} 
\smallskip
\caption{ Zeeman frequency data from 11 days of the Lorentz/CPT test (run 1)
  using the H maser.}
\label{f.Hdata}
\end{figure}

To identify any sidereal variations in $\nu_Z$, we fit a sidereal-period
sinusoid and a slowly varying background to the accumulated
$\nu_Z$ measurements. 
(See Fig.~\ref{f.Hdata} for the 11 days of data from run 1.)
Two coefficients, $\delta \nu_{Z,\alpha}$ and $\delta \nu_{Z,\beta}$, 
parameterize the sine and cosine components of the sidereal
oscillations.  ($\alpha$ and $\beta$ also correspond to non-rotating
directions in the plane perpendicular to the Earth's axis of
rotation.)
In addition, we used piecewise continuous linear terms (whose slopes
were allowed to vary independently for each day) to model the slow
drift of the Zeeman frequency.
In the field-inverted configuration, large variations in $\nu_Z$ led
to days for which this model did not successfully fit the data. Large
values of the reduced $\chi^2$ and systematic deviation of the
residuals from a normal distribution characterized such days, which
we cut from the data sample.
For each run, the fit determined the components $\delta
\nu_{Z,\alpha}$ and $\delta \nu_{Z,\beta}$ of the sidereal sinusoidal
variation (see Table I).
%
The total weighted means and uncertainties
for $\delta \nu_{Z,\alpha}$ and $\delta \nu_{Z,\beta}$ were then
formed from all three data sets, yielding the measured value
$A \equiv \sqrt{(\delta \nu_{Z,\alpha})^2 + (\delta \nu_{Z,\beta})^2} = 0.49
\, \pm \, 0.34$ mHz
(1-$\sigma$ level). This result is consistent with no observed
sidereal variation in the hydrogen $F=1$, $m_F= \pm 1$ Zeeman
frequency, given reasonable assumptions about the probability
distribution for $A$ \cite{A_distribution}.

Systematic sidereal-period fluctuations of $\nu_Z$ were smaller than
the 0.34 mHz statistical resolution.
The current in the main solenoid
typically varied by less than 5 nA out of 100 $\mu$A over 10 days,
corresponding to a change in $\nu_Z$ of $\sim 50$ mHz.  We corrected
the measured Zeeman frequency for this solenoid current drift.  The
sidereal component of the current correction was typically 
$25 \, \pm \, 10$
pA, corresponding to a sidereal-period variation of $\nu_Z \approx
0.16 \, \pm \, 0.08$ mHz.
The temperature inside the maser cabinet enclosure had a sidereal
component below 0.5 mK, corresponding to a sidereal-period modulation
of $\nu_Z$ of less than 0.1 mHz.
Potential Lorentz-violating effects acting directly on the electron
spins in the fluxgate magnetometer's ferromagnetic core could change
the field measured by the magnetometer and mask a potential signal
from the H maser experiment.  However, any such effect would be
greatly suppressed by a factor of 
%
%
%
\begin{minipage}[t]{\hsize}{
\small \begin{center} %
\begin{tabular}{crcrr}
\hline \hline
Run & Useful days & Field &  \multicolumn{1}{c}{$\delta\nu_{Z,\alpha}$} &
\multicolumn{1}{c}{$\delta\nu_{Z,\beta}$} \\
    & (cut days) & direction & \multicolumn{1}{c}{(mHz)} &
    \multicolumn{1}{c}{(mHz)} \\
\hline
    1 & 11 (0) & $\Uparrow$ & $0.43 \pm 0.36$ & $-0.21 \pm 0.36 $ \\ 
    2 & 3  (6) & $\Downarrow$ & $-2.02 \pm 1.27$ & $-2.75 \pm 1.41$ \\
    3 & 5  (7) & $\Downarrow$ & $4.30 \pm 1.86$ & $1.70 \pm 1.94$ \\
\hline \hline \\
\end{tabular}   
\end{center} %
\vskip -2pt %
{Table I. Means and standard errors for $\delta \nu_{Z,\alpha}$
  and $\delta \nu_{Z,\beta}$, the quadrature amplitudes of
  sidereal-period variations in the hydrogen $F=1$, $m_F = \pm 1$
  Zeeman frequency.  Results are displayed for each of three
  data-taking runs, listing also the number of days of useful data,
  the number of discarded data-taking days (in parentheses), and the
  direction of the maser's internal magnetic field in the laboratory
  frame.}
\\ \mbox{}
}%
\end{minipage}
%
$E/kT \sim 10^{-16}$ below the $\lesssim 1$ nG sensitivity of the
magnetometer, where $E$ is the Lorentz-violating shift of the electron
spin energy (known to be $\lesssim 10^{-29}$ GeV \cite{adelharris})
and $T$ is the temperature of the spins when the core is in zero
magnetic field (the equilibrium condition of the magnetometer lock
loop).  Also, the magnetic shielding reduces field fluctuations at the
magnetometer by a factor of only 6 whereas fluctuations at the
storage-bulb are reduced by 32,000.  Therefore, any effective magnetic
field shifts induced in the magnetometer by Lorentz/CPT-violations
were negligible in the present experiment.
Spin-exchange collisions between the H atoms shift the zero crossing
of the double resonance from the true Zeeman frequency \cite{savard}.
Hence, the measured $\nu_Z$ varies with H density in the maser.  We
monitored the atomic density by measuring the output maser power, with
the relation to $\nu_Z$ being $\lesssim 0.8$ mHz/fW.  During long term
operation, the average maser power drifted less than 1 fW per day.
The sidereal component was typically less than 0.05 fW, corresponding
to a 0.04 mHz variation in the Zeeman frequency.
Combining these systematic errors in quadrature with the statistical
uncertainty produces a final limit on a sidereal variation in the
hydrogen $F=1$, $\Delta m_F = \pm 1$ Zeeman frequency of 0.37 mHz,
which expressed in energy units is $1.5 \times 10^{-27}$ GeV.

The hydrogen atom is directly sensitive to Lorentz and CPT violations
of the proton and the electron. Following the notation of
Refs.~\cite{clockcompare,bkrhydrogen}, one finds that a limit on a
sidereal-period modulation of the Zeeman frequency ($\delta \nu_Z$)
provides a bound on the following parameters in the standard model
extension of Kosteleck\'y and co-workers:
\begin{equation}
\left| 
\tilde{b}^p_3 + \tilde{b}^e_3 \right| \le 2 \pi \delta \nu_Z
\label{e.BKRshift}
\end{equation}
for the low static magnetic fields at which we operate. (Here we have
taken $\hbar=c=1$.)  The subscript 3 in Eq.~(\ref{e.BKRshift})
indicates the direction along the quantization axis of the apparatus,
which is vertical in the lab frame.  The superscripts $e$ and $p$
refer to the electron and proton, respectively.

As in Refs.\ \cite{clockcompare,deh99}, we can re-express the time
varying change of the hydrogen Zeeman frequency in terms of parameters
expressed in a non-rotating inertial frame as
\begin{equation}
2 \pi \delta \nu_{Z,J} = \left( \tilde{b}^p_J + \tilde{b}^e_J \right)
\sin{\chi},
\label{e.Hnonrot}
\end{equation}
where $J$ refers to either of two orthogonal directions perpendicular
to the earth's rotation axis and $\chi=48^\circ$ is the co-latitude of
the experiment.

As noted above, a re-analysis of existing data from a spin-polarized
torsion pendulum \cite{adelharris} sets the most stringent bound to
date on Lorentz and CPT violation of the electron: $\tilde{b}^e_J
\lesssim 10^{-29}$ GeV \cite{hec99}.  Therefore, the H maser
measurement reported here constrains Lorentz and CPT violations of the
proton: $\tilde{b}^p_J \le 2 \times 10^{-27}$ GeV at the one sigma
level. This limit is comparable to that derived \cite{clockcompare}
from the ${}^{199}$Hg/${}^{133}$Cs clock comparison experiment of
Hunter, Lamoreaux \emph{et al.} \cite{hl9x} but in a much cleaner
system: the hydrogen atom nucleus is simply a proton, whereas
significant nuclear model uncertainties affect the interpretation of
experiments on many-nucleon systems such as ${}^{199}$Hg and
${}^{133}$Cs.

To our knowledge, no search for sidereal variations in the hydrogen
Zeeman frequency has been performed previously. Nevertheless, implicit
limits can be set from a widely-practiced H maser characterization
procedure in which the Zeeman frequency is measured by observing the
drop in maser output power induced by a drive field swept through the
Zeeman resonance \cite{kgrvv,va}.  It is reasonable to assume that
sidereal-period variations of the Zeeman frequency of $\sim1$ Hz would
have been noticed.  Thus, our result improves upon existing implicit
constraints by over two orders of magnitude.

In conclusion, precision comparisons of atomic clocks provide
sensitive tests of Lorentz and CPT symmetries
\cite{clockcompare,spontlorentz,stringLorentz,kosanalysis,bkrhydrogen}.
A new measurement with an atomic hydrogen maser provides a clean limit
on Lorentz and CPT violation involving the proton that is consistent
with no effect at the $10^{-27}$ GeV level. Further details of this
work will be found in Ref.~\cite{longLorentz}.

We gratefully acknowledge the encouragement and assistance of Alan
Kosteleck\'y.  Financial support was provided by NASA grant NAG8-1434
and ONR grant N00014-99-1-0501. M.\ A.\ H.\ acknowledges a fellowship
from the NASA Graduate Student Researchers Program.

\end{multicols} 

\begin{thebibliography}{xx}
\bibitem{zeemanAsym} 
At 0.6 mG the differential splitting of the $m_F=+1$ and $-1$ 
levels is $<1$ mHz, and is included in the fit model for $\nu_Z$.
\bibitem{A_distribution} 
In the case where $\delta\nu_{Z,\alpha}$ and $\delta\nu_{Z,\beta}$ have zero 
mean value and the same variance $\sigma$, the probability 
distribution for $A$ takes the form 
$P(A) = \sigma^{-2}A \exp(-A^{2}/2\sigma^{2})$, with the most 
probable value of $A$ occurring at $A = \sigma$.
\end{thebibliography}
\end{document}